# Enhanced Emission and Two-Photon Interference of Lead-Vacancy Centers in Diamond Solid Immersion Lenses

Koyo Hirai [1,†], Eiki Ota [1,†], Yiyang Chen [1], Ren Kato [1], Peng Wang [1,‡], Toshiharu Makino [2], Takashi Taniguchi [3], Masashi Miyakawa [3], Shinobu Onoda [4], Mutsuko Hatano [1], and Takayuki Iwasaki [1,*]

[1]*Department of Electrical and Electronic Engineering, School of Engineering, Institute of Science Tokyo, Meguro, 152-8552 Tokyo, Japan*

[2]*Advanced Power Electronics Research Center, National Institute of Advanced Industrial Science and Technology, Tsukuba, 305-8568 Ibaraki, Japan*

[3]*Research Center for Materials Nanoarchitectonics, National Institute for Materials Science, Tsukuba, 305-0044 Ibaraki, Japan*

[4]*Takasaki Advanced Radiation Research Institute, National Institutes for Quantum Science and Technology, Takasaki, 370-1292 Gunma, Japan*

*Corresponding author: Takayuki Iwasaki (iwasaki.t.c5b4@m.isct.ac.jp).

[†]These authors contributed equally to this work.

[‡]Present address of Peng Wang: Institute for Quantum Optics, Ulm University, Albert-Einstein-Allee 11, D-89081 Ulm, Germany.

## Abstract

A negatively charged lead-vacancy ($PbV^-$) center in diamond is a novel quantum system which can be operated at a high temperature above 4 K owing to its large ground state splitting. To unlock the quantum properties of the $PbV^-$ center, the enhancement of the fluorescence intensity is a key issue. Here, we demonstrate enhanced emission from $PbV^-$ centers by a factor of approximately 10 using solid immersion lenses (SILs) fabricated on rough diamond surface caused by high-temperature anneal over 2000°C. Resonant excitation reveals the narrow emission close to the transform-limited linewidth. Furthermore, we demonstrate two-photon interference using one single $PbV^-$ center in an SIL. This enhanced emission and indistinguishability will lead to further development of the $PbV^-$ center towards quantum network node applications.

## 1. Introduction

A bright quantum emitter is expected to serve as nodes for high-rate large-scale quantum networks [1,2]. Color centers in solid-state materials have emerged as promising quantum emitters due to their optical and spin properties and ease of optical access and readout. Among them, negatively charged nitrogen-vacancy ($NV^-$) centers in diamond have been extensively studied, including such as two-photon interference [3,4], single-shot readout [5], heralded entanglement [6], and a three-node quantum network [7]. However, spectral diffusion [8] of $NV^-$ centers originating from the $C_{3v}$ symmetry remains a significant issue, and is considered to pose a challenge for high-rate quantum communication. In recent years, group-IV vacancy color centers with the $D_{3d}$ symmetry [9] have attracted increasing attention. This is because their inversion symmetry makes them robust against external electric-field noise and the Debye-Waller factor is higher than that of $NV^-$ center [10–13]. To date, the split vacancy configuration of negatively charged silicon-vacancy ($SiV^-$) [14–17], germanium-vacancy ($GeV^-$) [18–20], tin-vacancy ($SnV^-$) [21–26], and lead-vacancy ($PbV^-$) [27–30] centers has been revealed. Particularly, $PbV^-$ are anticipated to show a long spin coherence time at higher temperatures well above 4 K since it is highly resistant to phonon interactions due to the large ground-state splitting ($\Delta_{GS}$ $\approx$ 3900 GHz [30]). Our previous studies have shown that $PbV^-$ can be formed by combining Pb ion implantation with high-pressure and high-temperature (HPHT) treatment over 2000℃, achieving photon emission with nearly transform-limited linewidths even above 10 K [31] and narrow inhomogeneous distributions of the central wavelengths (standard deviations of ~5 GHz) [32].

To further accelerate the evaluation of the optical and spin properties of the $PbV^-$ center, it is significant to embed the emitter into a diamond photonic structure which can enhance the photon extraction [33]. Efficient photon extraction from color centers except the $PbV^-$ center in diamond has been reported by fabricating various structures such as nanopillar [23,34–36], solid immersion lens (SIL) [37–39], nanophotonic cavity [26,40–43], and waveguide [44–46]. While the high-quality $PbV^-$ centers can be fabricated by the HPHT treatment, this process considerably degrades the surface roughness of the diamond substrates, due to etching and regrowth of diamond [30]. Therefore, realizing the formation of high-quality $PbV^-$ centers in efficient photon collection structures remains a challenging task.

In this study, we demonstrate the fabrication of the diamond SILs incorporating $PbV^-$ centers on rough HPHT-treated diamond substrates and the enhancement of the fluorescence intensity from the $PbV^-$ centers. We obtain approximately 10 times higher fluorescence intensity under non-resonant excitation and nearly transform-limited linewidth under resonant excitation. Finally, we observe the indistinguishability of photons through Hong-Ou-Mandel (HOM) interference [47] using one single $PbV^{-}$ in SIL.

## 2. Results and Discussion

### 2. 1. SILs fabricated on HPHT-treated diamond substrates

A $PbV^{-}$ center takes a split-vacancy configuration, in which a Pb atom is at an interstitial position between two vacancies (Fig. 1(a)). To incorporate high-quality $PbV^{-}$ centers, we fabricate SILs on samples with rough surface formed during the high-temperature anneal over 2000℃ under high pressure (Fig. 1(a)). Mainly, we use two samples used in this study with different fabrication conditions (see Experimental Section for details). They have the mean roughness of 20.1 nm (Sample 1) and 83.9 nm (Sample 2), as shown in Fig. 1(b) (also see Section S1 for details of the measurement regions). The high roughness is thought to be caused by etching and growth during the HPHT anneal process [30,48]. Even on the rough surface, we can fabricate micro-scale SILs. Figure 1(c) shows an SEM image of a 63 SIL array in Sample 1, taken immediately after the focused ion beam (FIB) milling. We can see roughening of the diamond surface in the surrounding area. In a magnified image of one SIL (Fig. 1(d)), the height and radius of the SIL are estimated to be approximately 500 nm and 550 nm, respectively. Note that the bright white objects on top of the SIL and untreated diamond are the Pt/Pd coating layer for the FIB process. The sidewall and trench regions of the SIL, which are exposed to Ga ions during the fabrication, look very smooth. This is likely attributed to a FIB-induced planarization effect. A previous study has reported that Ga-ion-beam irradiation during FIB processing forms an amorphous carbon surface layer on diamond surfaces [49,50]. In addition, the recoil of atoms in the host occurs by ion collision cascades. Thus, host atoms are locally redistributed, resulting in smoothing of the solid surface [51]. Interestingly, although the ion-beam irradiation area is defined as a concentric-ring pattern (see 4. Experimental Section), the diameter of the fabricated structures decreases from the bottom toward the top, resulting in the hemispheric-like structures. This phenomenon is attributed mainly to stray-dose etching, charge accumulation, and transverse ion scattering, which is reported in a previous study [52]. The top area of the SIL appears slightly rough, and furthermore, there remains a small amount of the Pt/Pd layer on top. Thus, the rough surface morphology after HPHT annealing remains to some extent at the top region. Figure 1(e) shows an atomic force microscope (AFM) image after removal of the Pt/Pd layer. The middle panel shows a line profile of the AFM image. Although the diamond surface is not perfectly flat, the SIL structure is clearly observed, suggesting that a perfectly smooth diamond surface is not a prerequisite for SIL fabrication. The magnified image of the SIL (Fig.

1(e), bottom) further reveals the rounded geometry of the fabricated SIL with smooth sidewall and almost flat trench.

Figures 1(f, g) show SEM and AFM images of a SIL fabricated on the rougher diamond surface (Sample 2), acquired after removing the Pt/Pd layer. Although the surface of Sample 2 exhibits a larger roughness due to the higher temperature and longer time annealing, hemispheric-like structures are fabricated even on this surface. We find that this SIL becomes an elliptical shape, probably caused by the high roughness and FIB conditions including the focus adjustment (see Section S2). Even with the presence of larger surface irregularities (Fig. 1(g), middle), SIL is fabricated and the trench part around the SIL becomes smooth (Fig. 1(g), bottom). It is worth noting that in another sample, we find that the fabricated SILs are severely distorted on an area where large diamond pyramid-like structures grow during the HPHT annealing (see Section S3 for details).

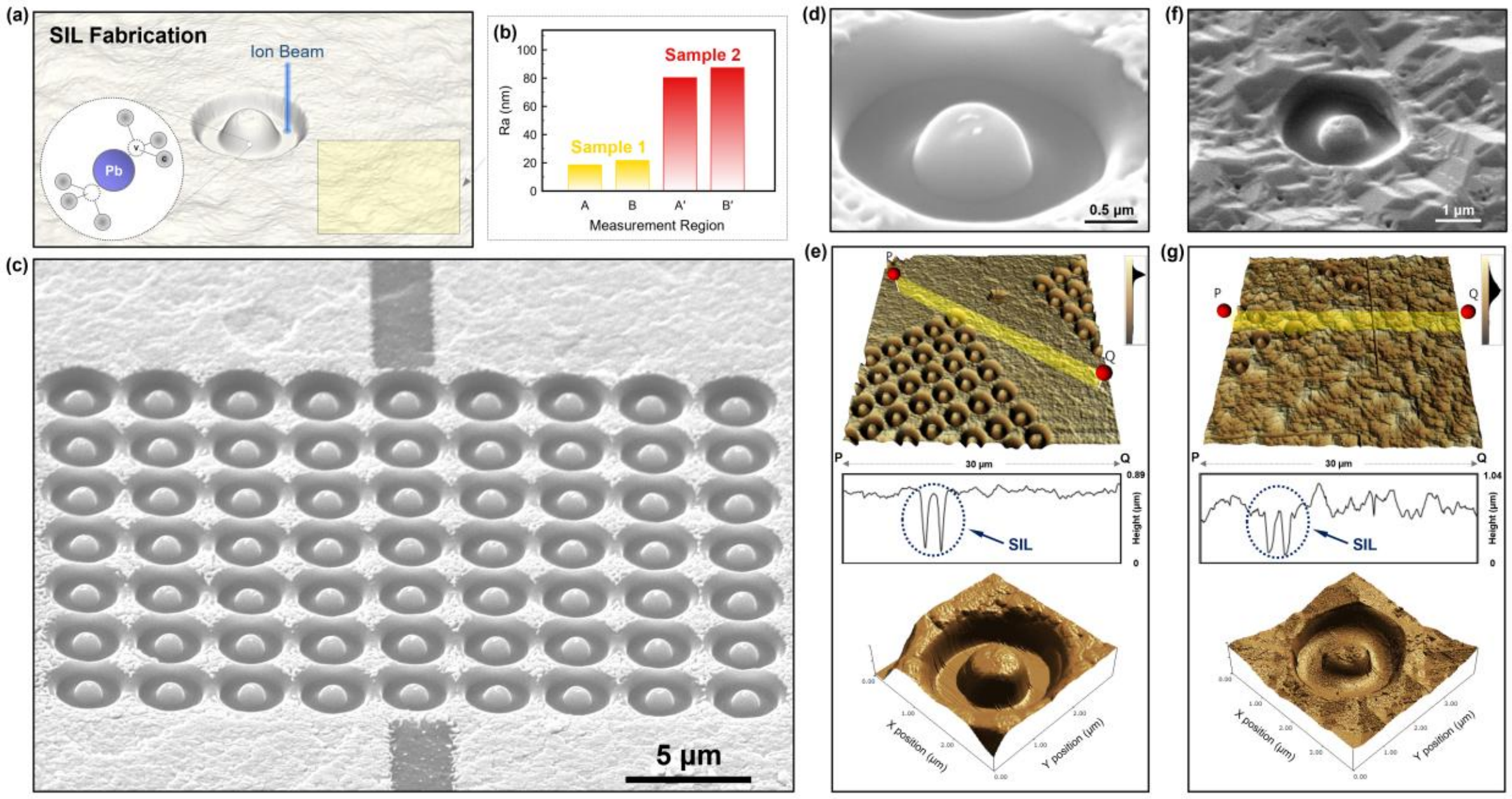

**Figure 1 | Diamond SILs fabricated on HPHT-treated rough diamond surface.** (a) Schematic illustration of a SIL incorporating a $PbV^{-}$ center. (b) Ra roughness of the two samples. Two regions near SILs are measured for each sample. (c) SEM image of a fabricated SIL array (7 × 9) in Sample 1 with a Pt/Pd coating layer. (d) Magnified SEM image of an SIL within the array in (a). (e) AFM observation of Sample 1. Top: wide area scan (30 × 30 $\mu m^2$), middle: line profile indicated in top panel, bottom: high-resolution scan (3 × 3 $\mu m^2$). (f) SEM image of an SIL in Sample 2. (g) AFM observation of Sample 2. Top: wide area scan (30 × 30 $\mu m^2$), middle: line profile indicated in top panel, bottom: high-resolution scan (4 × 4 $\mu m^2$). The SEM images are acquired at a tilt of 52°.

## 2. 2. Enhanced fluorescence from $PbV^-$ centers embedded in SILs

We observe optical properties of $PbV^-$ centers embedded in SILs. In sections 2.2 and 2.3, we use Sample 1 for optical characterizations, unless otherwise specified. Sample 1 contains an ensemble of $PbV^-$ centers (see Section S4 for details), enabling us to efficiently incorporate multiple $PbV^-$ centers into the SILs. Figure 2(a) depicts a confocal fluorescence microscope (CFM) mapping of an SIL array, in which some of the SILs apparently show strong fluorescence. We mainly measure SIL-1, marked in red in Fig. 2(a) for further characterizations. Figure 2(b) shows a low-temperature PL spectrum. Two peaks at ~550 nm and ~554 nm correspond to C- and D-transitions of the $PbV^-$ center, respectively, shown in the inset of Fig. 2(b), which agree with previous studies [30–32]. Then, we perform a PLE scan, showing one strong peak at a wavelength of ~550.5286 nm and small peaks between 550.534 nm and 550.548 nm (Fig. 2(c)). Therefore, multiple $PbV^-$ centers are embedded in the SIL as expected. The largest peak occupies 85.6% of the total fluorescence in the spectrum. The differences in the fluorescence enhancement among the peaks are considered to be caused by the position of the emitters in the SIL. A field of view diameter, $d_{FOV}$, of SIL is described as follows [39,53,54]:

$$d_{FOV} = \sqrt{\frac{2r\lambda}{n(n-1)}},$$

where λ is the emission wavelength of a color center, n is the refractive index of the host material, and r is the radius of the SIL. A diamond SIL with a radius of 550 nm leads to $d_{FOV}$ ~420 nm for ZPL of the $PbV^-$ center (550 nm). Therefore, an emitter shifted more than 210 nm from the center of SIL should show less fluorescence intensity, as seen in Fig. 2(c).

Interestingly, even with the ensemble state, we find that SIL-1 demonstrates the single photon nature in Hanbury-Brown and Twiss (HBT) measurement [55] (Fig. 2(d)). The second-order autocorrelation function at zero time delay, $g^2(0)$, becomes clearly less than 0.5. The dominant large peak with the highest count rate in the PLE scan (Fig. 2(c)) is thought to behave as a single photon emitter, while others become background to increase $g^2(0)$. Indeed, when we observe two comparable PLE peaks in another SIL, $g^2(0)$ goes to nearly 0.5, corresponding to double emitters [56] (see Section S5 for details).

Figure 2(e) shows a saturation curve of $PbV^-$ centers in SIL-1 (red). For comparison, we use a single emitter on a bulk region without SILs in Sample 2 fabricated with a lower Pb fluence fabrication (blue). Note that the bulk emitter is characterized after the first 20 min anneal (see Experimental Section and Section S6 for details). The fluorescence intensity increases with the 532 nm laser power and shows saturation. The data are fitted with an equation [15,30]:

$$I = I_\infty \times \frac{P}{P + P_{\mathrm{sat}}},$$

where, $I_\infty$ and $P_{\mathrm{sat}}$ are saturation intensity and saturation power, respectively. The saturation power of SIL-1 and Bulk are ~1.2 Mcps and ~104 kcps, respectively. We consider two background components: the small peaks observed in the PLE spectrum (Fig. 2(c)) and background of an empty SIL. The background-subtracted saturation curve is depicted in Fig. 2(e) (Corrected SIL-1, orange). The saturation intensity of the corrected SIL-1 is ~1 Mcps. Similar to SIL-1, the background of Bulk is subtracted by the saturation curve of an empty SIL. Thus, compared with the emitter in bulk, we obtain approximately 10 times enhancement of the fluorescence intensity in the SIL. The enhanced count rate is also confirmed in PLE spectrum. A high peak count of 219 kcps is obtained even under a weak resonant laser power of 0.3 nW in SIL (Fig. 2(c)), while the count in Bulk is limited to 37 kcps with 1.1 nW resonant laser (see Section S6). Several studies have reported the fabrication of SILs on diamond surfaces using FIB milling for $NV^-$ centers [54,57–61], $SiV^-$ centers [16,37,62], $GeV^-$ centers [20,38,39] and $SnV^-$ centers [63]. Most reported enhancement factors are in the range of 3–10 [37,39,54,57–59,61]. Therefore, our result is considered reasonable and consistent with previous studies.

The saturation powers of SIL-1 and Bulk are ~1.44 mW and ~3.66 mW respectively. The fluorescence of SIL-1 is saturated at less than half the excitation power required for the emitter in bulk. A similar tendency has been reported for $NV^-$ centers in SILs [57]. The reduced saturation power is attributed to the increased effective numerical aperture of the SIL, which reduces the laser excitation volume and boosts the effective power density for a given input laser power [57,64].

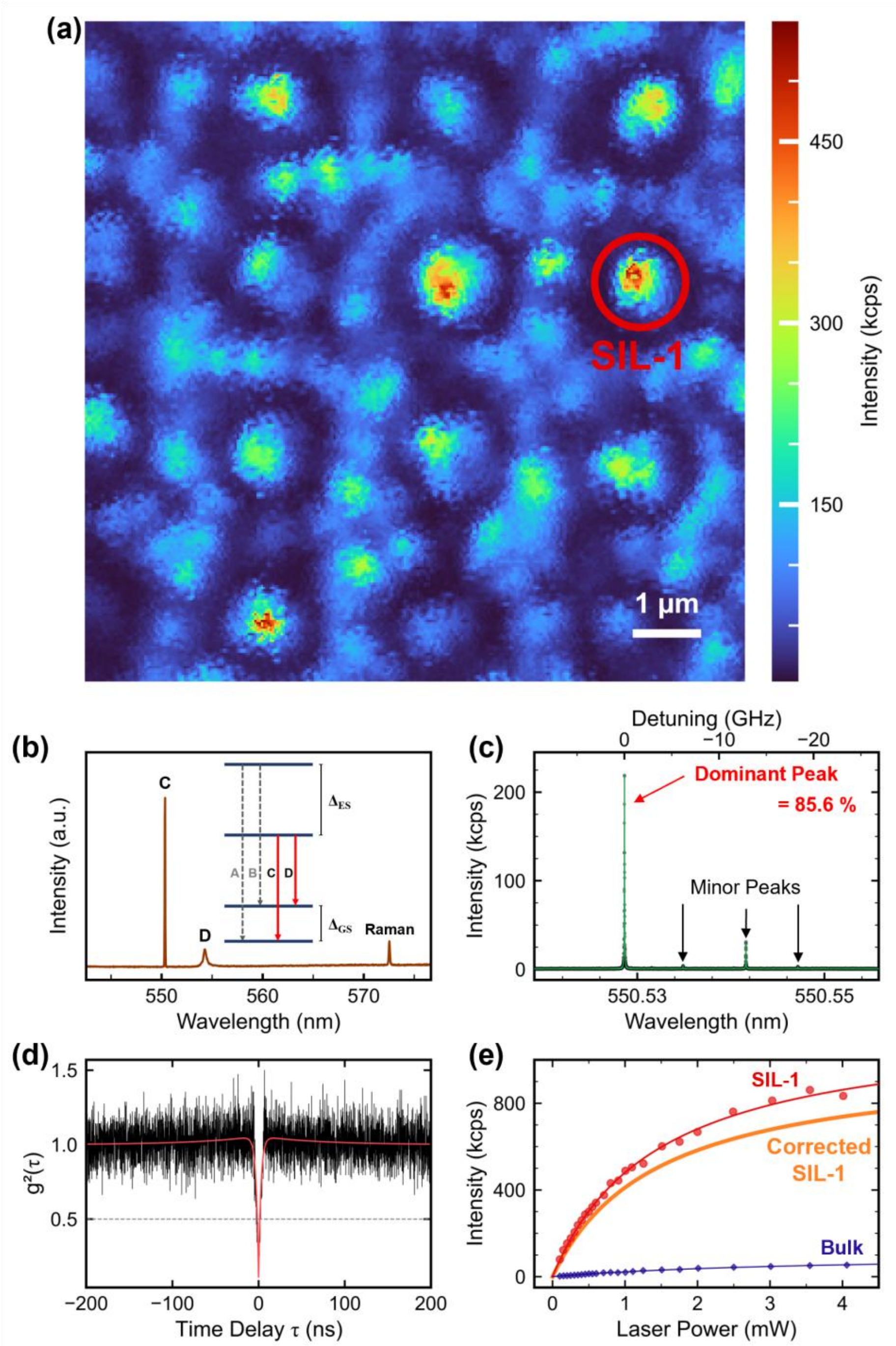

**Figure 2 | Enhanced photon emission from PbV⁻ centers embedded in SIL.** (a) CFM image of PbV⁻ fluorescence in SILs. (b) PL spectrum of the PbV⁻ centers at 6.4 K. Inset: Energy-level diagram of the PbV⁻ center. (c) Wide frequency range PLE scan of the C-transition from PbV⁻ centers. (d) Second-order autocorrelation function, $g^2(\tau)$. The HBT curve is fitted with an equation [23,30] of $g^2(\tau) = 1 - c\{(1+b)\exp[-|\tau|/\tau_1] - b\exp[-|\tau|/\tau_2]\}$, where $b$, $c$, $\tau_1$, and $\tau_2$ are fitting parameters. (e) Saturation curves of ZPLs from SIL-1 and Bulk. In the panels (a, d, e), The ZPLs of PbV⁻ centers are detected with a 550 nm band pass filter (BPF), thus including both C- and D-peaks. For the PLE scan in (c), phonon sideband (PSB) of PbV⁻ centers is acquired through a 561 nm long pass filter.

## 2. 3. Transform-limited photon emission from $PbV^-$ centers in SILs

We observe a narrow frequency range PLE spectrum of the enhanced $PbV^-$ center to see the optical coherence. Figure 3(a) shows a PLE spectrum of the C-transition from the dominant $PbV^-$ center in SIL-1. A linewidth of the PLE peak is 41.9 MHz, which is comparable to a previous report [31]. From a lifetime measurement using a 532 nm pulsed laser in Fig. 3(b), we estimate a lifetime $\tau$ of 4.3 ns, corresponding to the transform-limited linewidth of 37 MHz. Thus, a nearly transform-limited emission can be obtained from the $PbV^-$ center even in the SIL. Note that the clear PLE spectrum can be observed after the SIL is exposed to an ICP process [65] that etches the surface of 30 nm. Without the ICP process, a PLE spectrum is only observed with simultaneous 532 nm non-resonant laser irradiation to stabilize the charge state [66] (see Section S7 for more details). We attribute this phenomenon to the surface damage induced by Ga ions during FIB milling, which introduces an excessive number of defects and influences the charge state of $PbV^-$ centers.

Then, we investigate PLE spectra from multiple enhanced $PbV^-$ centers across multiple SIL arrays. The inhomogeneous and linewidth distributions of 20 emitters are summarized in Figs. 3(c, d) (see Section S8 for each spectrum). A Gaussian fit to the inhomogeneous distribution yields a width of 11.5 GHz, which is comparable to the bulk sample in previous work [32]. From the linewidth distribution (Fig. 3(d)), the mean linewidth (μ) is 40.6 MHz, which is slightly larger than the average transform-limited linewidth estimated from Figs. 3(b, g). It is worth noting that we find a few emitters showing charge state transfer during the measurements and that they are not included in the statistical analyses of the PLE spectra since the linewidths cannot be reliably determined (see Section S5 for more details).

Figure 3(e) shows the stability of PLE peak of the dominant peak in SIL-1. No spectral diffusion is observed for more than 20 min, indicating that our SIL can host the optically stable $PbV^-$ while enhancing the fluorescence collection efficiency.

To examine temperature robustness of a $PbV^-$ center in an SIL, the temperature dependence of the PLE linewidth is examined on a $PbV^-$ center (Figs. 3(f, g)). This emitter shows a linewidth of 41.0 MHz at 6.5 K, close to the transform-limited linewidth of 35 MHz estimated from an inset of Fig. 3(g). The narrow linewidth remains up to ~16 K, while it broadens at higher temperatures, as observed on a $PbV^-$ center in bulk diamond [31]. The observation of the transform-limit even above 10 K is owing to suppression of the phonon interaction of the $PbV^-$ center with a large ground state splitting. Note that the intensity drops on high temperatures are attributed to vibration of the cryostat.

Here, we confirm that the $PbV^{-}$ centers in SILs show nearly transform-limited linewidth, long-time stability, narrow inhomogeneous distribution, and temperature robustness, comparable to emitters in bulk diamond.

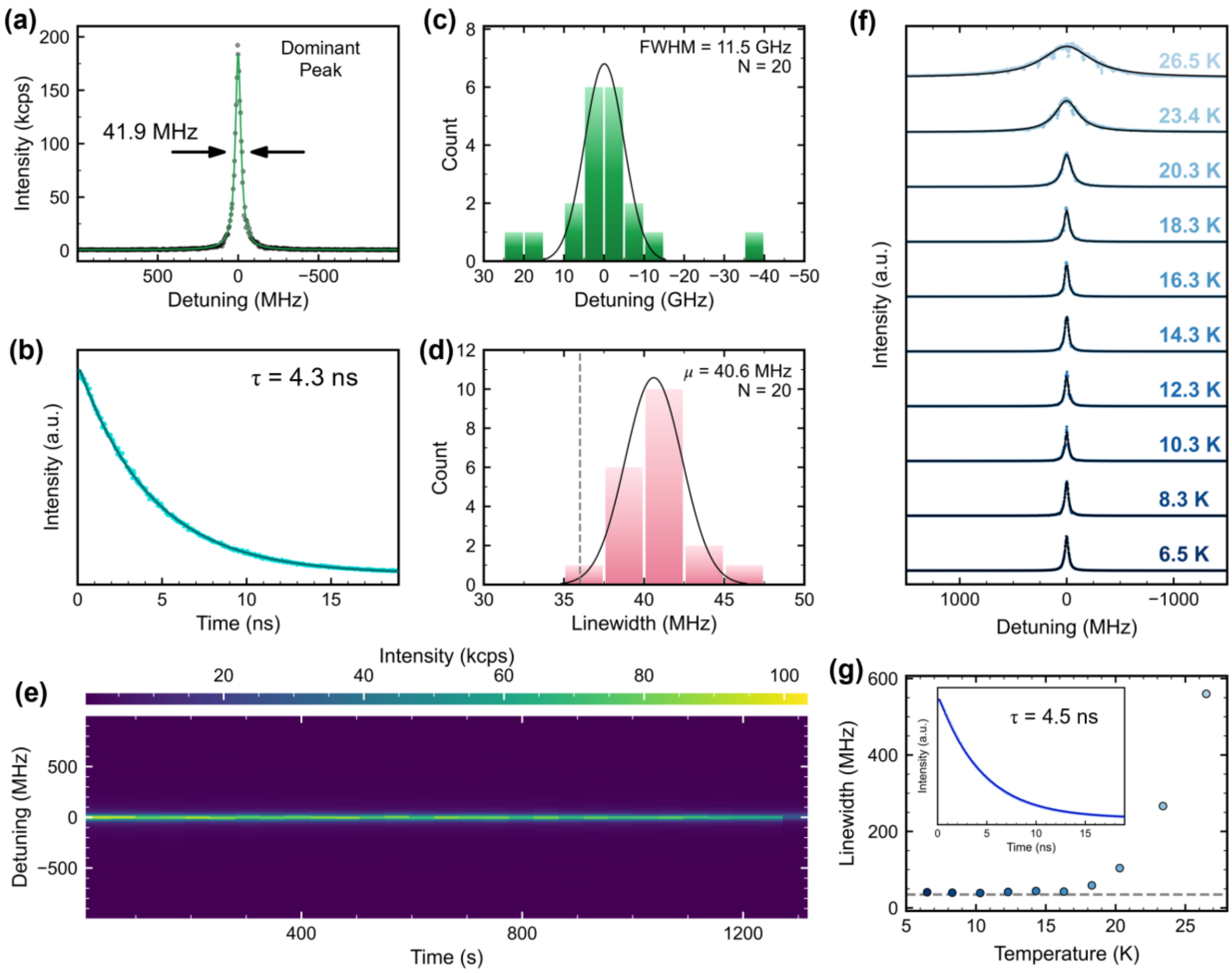

**Figure 3 | Nealy transform-limited emission in SIL.** (a) PLE scan of the dominant C-peak in SIL-1. The resonant excitation power is ~0.3 nW. The center wavelength of the peak is 550.528636 nm. (b) Lifetime measurement of the same emitter using 532 nm pulse laser. (c) Inhomogeneous distribution of fluorescence-enhanced $PbV^{-}$ centers in SILs. A detuning of 0 GHz corresponds to a wavelength of 550.53 nm. (d) Linewidth distribution. The dashed line indicates the transform-limited linewidth of 36 MHz calculated from the average lifetime from two lifetime data. The data in (c, d) are fitted by a Gaussian function. (e) Stability of the PLE peak under the 0.1 nW resonant excitation power. The low laser power is used to prevent transfer to the dark charge state. (f) PLE spectra at different temperatures. The resonant excitation power during this measurement is 0.2–0.3 nW. The centers of the peaks are aligned. (g) Temperature-dependent linewidth. The dashed line indicates the transform-limited linewidth of 35 MHz calculated from the lifetime (4.5 ns) shown in inset.

### 2. 4. Two-photon interference using a single $PbV^-$ center in SIL

Finally, we perform HOM interference with one single $PbV^-$ center in a SIL. Here, Sample 2 with a lower Pb density is used to obtain a single emitter in SIL. Because of misaligning the locations of the emitter and SIL fabrication, we find an emitter in one of seven SILs (Fig. 4(a)). The fluorescence enhancement from the target $PbV^-$ center is estimated to be by a factor of approximately 4. Figure 4(a) also shows the experimental setup for the HOM interference. PL and PLE spectra of the $PbV^-$ center are measured using a spectrometer and APDs after a pinhole. For the interference, photons emitted from the $PbV^-$ center excited using non-resonant 532 nm laser are transferred to the interference system.

The C-peak of this emitter shows a transform-limited linewidth of 36 MHz under resonant excitation measured with PSB through a 561 nm LPF (Fig. 4(b)). For the interference, we remove the D-peak of the $PbV^-$ center by using a 554 nm short-pass filter (SPF) in addition with a 550 nm BPF, as shown in Fig. 4(c). The C-peak photons are delivered with a 5 m single mode fiber and are split into two paths at the first beam splitter (BS). One of the two paths is delayed by 9 m ($\Delta\tau \sim 44$ ns) using a longer single-mode fiber. Then, the photons enter into the second beam splitter for the interference. The polarization of the photons is controlled with polarization optics before the second BS.

First, we perform the HBT measurement by blocking one input arm of the second BS (Fig. 4(d)), effectively configuring the setup as a HBT interferometer [67]. By fitting the measured data considering the instrument response function (IRF) associated with the setup timing jitter [68], we obtain a zero-delay correlation of $g^2(0)$ = 0.21. After deconvolving the IRF, the intrinsic single-photon purity improves to $g^2(0)$ = 0.02, confirming that the emitter operates as a pure single-photon source [69].

Then, HOM experiments are conducted by changing the polarization of the photons just before the second BS. In the case of the orthogonal polarization between photons from the two paths, the zero-delay correlation $g_\perp(0)$ clearly exceeds 0.5, as expected for distinguishable photons (orange curve in Fig. 4(e)). In contrast, the value drops below 0.5 for the parallel polarization (blue curve in Fig. 4(e)). The second-order correlation function for the HOM curve is fitted with the following expression [70]:

$$g^2_{\mathrm{HOM}}(\tau) = 4(T_A^2 + R_A^2)R_B T_B g^{(2)}(\tau) + 4R_A T_A T_B^2 g^{(2)}(\tau - \Delta\tau) + 4R_A T_A R_B^2 g^{(2)}(\tau + \Delta\tau) \\ - 4R_A T_A (T_B^2 + R_B^2)\eta \left|g^{(1)}(\tau)\right|^2$$

where the coefficients $R_A (R_B)$ and $T_A$ $(T_B)$ denote the reflectance and transmittance of the first (second) BS located at the input (output) stage of the Mach–Zehnder interferometer, respectively. $\Delta\tau$ corresponds to the temporal delay introduced between the two interferometer arms (~44 ns). The factor $\eta$ accounts for the reduction in interference visibility due to environmental imperfections such as mode mismatch, decoherence, and other experimental inefficiencies [71]. $g^1(\tau)$ denotes the first-order correlation function, which characterizes its coherence property [70]. Figure 4(e) show HOM interference curves for the orthogonal and parallel polarizations (see Section S9 for wider range data). Fitting with the model above with considering the IRF, we extract $g_\perp(0) = 0.68$, $g_\parallel(0) = 0.36$ Thus, the visibility defined as $V = 1 - {g_\parallel(0)}/{g_\perp(0)}$ becomes ~47%. The deconvolution of the IRF enhances the HOM visibility to 62%. The remaining gap to the ideal limit can be primarily attributed to intrinsic emitter dynamics under non-resonant excitation. While the $PbV^-$ center basically exhibits robust charge-state stability due to the inversion symmetry, the non-resonant 532 nm continuous-wave excitation drives dynamic charge fluctuations in the surrounding defect [72–75] such as divacancies [76]. This phenomenon is thought to cause the line broadening compared with the PLE in Fig. 4(c) under resonant excitation, thereby limiting photon indistinguishability [38,77]. Potential optical setup imperfections such as residual spatial wavefront mismatch or polarization drift in the MZI may also cause a reduced observable contrast [71,78].

Visibilities for $NV^-$ [3,4,6], $SiV^-$ [17], $GeV^-$ [38] and $SnV^-$ [77,78] centers range from 35% to 80%. Our deconvoluted visibility of 62% indicates that the single $PbV^-$ center possesses a value in the similar level. A higher interference visibility of over 95% for $SnV^-$ has been recently reported with highly optimized resonant excitation and time filtering [71]. Adopting a similar experimental strategy will enable us to obtain comparable performance for $PbV^-$ centers.

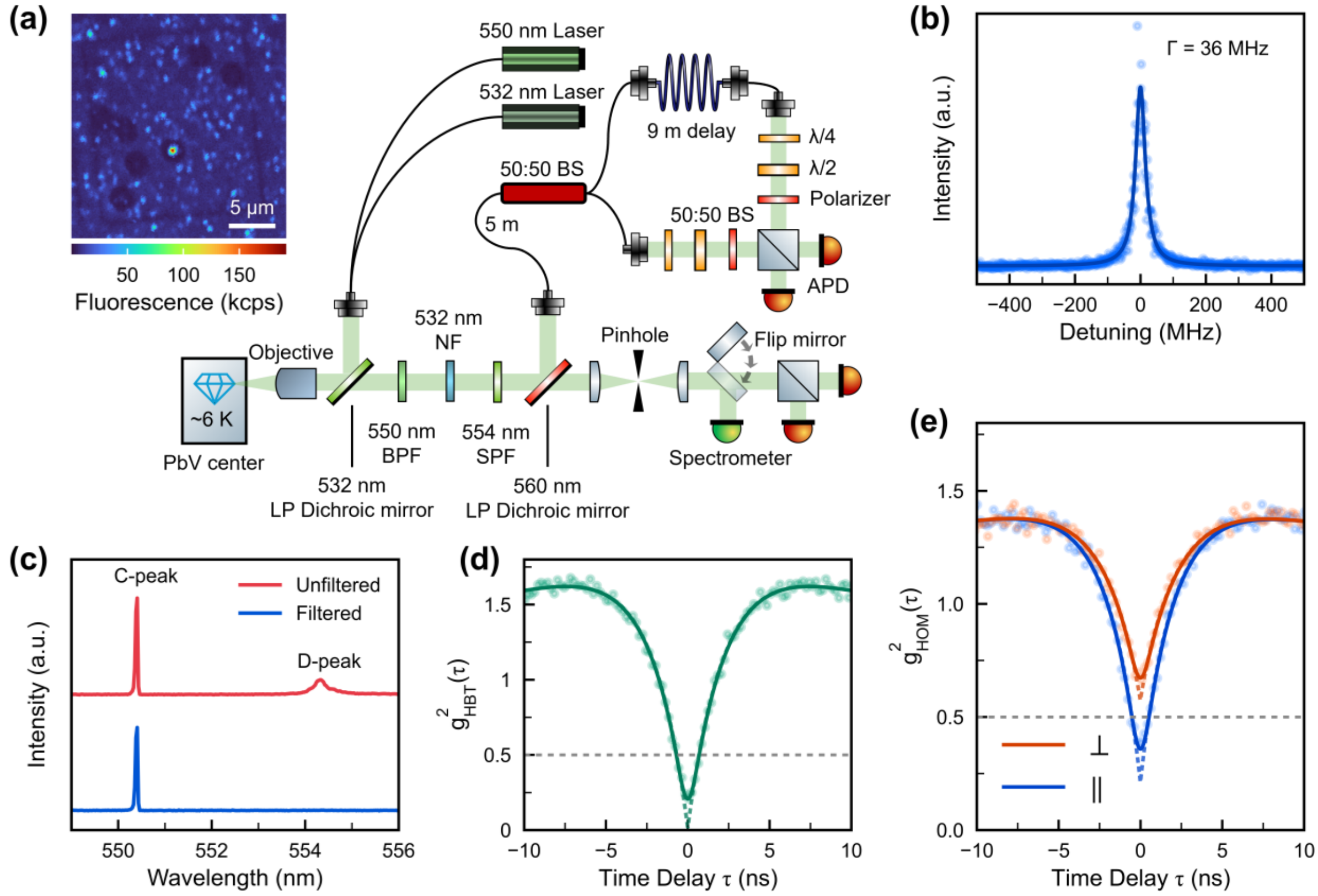

**Figure 4 | Two-photon interference from a single PbV⁻ center in SIL.** (a) The CFM image of the target PbV⁻ center in SIL and experimental setup for the HOM interference. Non-resonant 532 nm laser is employed for the interference measurements. (b) PLE spectrum with a transform-limited linewidth of 36 MHz. (c) PL spectrum without and with the 554 nm SPF, showing the selection of the C-transition. (d) HBT measurement and (e) HOM interference measurements. Solid lines show fits to the experimental data, incorporating a system timing jitter of 750 ps in the setup. Dashed lines show the deconvoluted curves.

## 3. Conclusion

We fabricated micro-size SILs on HPHT-treated rough diamond surfaces by FIB processing. We demonstrated about 10-fold fluorescence enhancement from the $PbV^-$ center embedded in SIL compared to the $PbV^-$ center in bulk diamond. Nearly transform-limited photon emission, narrow inhomogeneous distribution, and stability were confirmed, comparable to ones in bulk diamond. Using one $PbV^-$ center in SIL, we demonstrated the HOM interference with a visibility of 47%, increasing to 62% after deconvolution, which will be further improved by the timing jitter suppression and resonant-excitation-based interference without spectral diffusion. The developed structures with the enhanced fluorescence and transform-limited emission in this study are an important technology for characterization of the $PbV^-$ center, e.g. as used for spin control [79]. Thus, the advancement demonstrated in this paper will pave the way for further developments of the $PbV^-$ centers in diamond towards quantum network nodes.

## 4. Experimental Section

We implant Pb ions into IIa-type (001) single-crystal diamond substrates (Element Six Ltd., electronic grade) at an acceleration energy of 4.5 MeV. The fluence is $1\times10^8$ or $2\times10^9$ $cm^{-2}$. A projected depth of the Pb ions is approximately 500 nm according to the SRIM calculation [80] (see Section S11). After ion implantation, HPHT annealing [30,32] is performed at 2100–2300°C under a pressure of 7.7 GPa for 20–40 min, leading to the formation of high-quality $PbV^-$ centers. Annealing of Sample 2 is performed two times for 20 min. This is because PLE peaks are not likely to be observed on several spots after the first 20 min anneal step. Therefore, the additional second annealing is carried out. The fabrication conditions for two samples, denoted as Samples 1 and Sample 2, are summarized in Table 1. Sample 1 is used to characterize the morphologies and optical properties of $PbV^-$ centers in SIL (Figs. 1, 2, and 3). whereas Sample 2 is used to for the HOM experiments as well as the morphological characterization (Figs. 1 and 4).

**Table 1 | Sample fabrication conditions.**

| Sample | Pb ion fluence ($cm^{-2}$) | Acceleration energy (MeV) | Annealing temperature (°C) | Annealing pressure (GPa) | Annealing time (min) |
|---|---|---|---|---|---|
| Sample 1 | $2 \times 10^9$ | 4.5 | 2200 | 7.7 | 20 |
| Sample 2 | $1 \times 10^8$ | 4.5 | 1st step: 2100<br>2nd step: 2300 | 1st step: 7.7<br>2nd step: 7.7 | 1st step: 20<br>2nd step: 20 |

For the SIL fabrication, we employ a dual-beam FIB-SEM system (JIB-4500, JEOL) using Ga ions beam (30 keV). We create the SILs by drawing patters of concentric circles [57]. Prior to the fabrication, the diamond samples are covered with a ~20 nm thick Pt/Pd alloy layer to suppress charging of the diamond surface. After the fabrication of SILs, the Pt/Pd layer is removed with acid treatments. According to the SRIM simulation [80], Ga ions are distributed to a depth of approximately 30 nm from the surface. Thus, soft ICP etching [65] without applied bias is performed to clean the surface of the SILs. The samples are cleaned with hot piranha solution, a heated mixture of nitric acid and sulfuric acid before AFM and optical measurements. The morphologies of the diamond surface and SILs are observed using the FIB-SEM system and AFM (SPM-9700, Shimadzu).

The optical measurements of the $PbV^-$ centers are conducted utilizing a cryogenic home-built confocal microscope system. The samples are mounted in the cryostat and measured at 5-7 K over the entire experiments except the temperature dependence (Figs. 4(f,g)). We employ a 532 nm laser for non-resonant excitation. We observe the fluorescence of $PbV^-$ centers with either avalanche photodiodes (SPCM-AQRH-14, Excelitas) or a spectrometer (SpectraPro HRS-300, Teledyne Princeton Instruments) with a CCD camera (PIXIS 100BReX-U, Teledyne Princeton Instruments). We employ a tunable dye laser (Matisse 2 DS, Sirah Lasertechnik) or a tunable diode laser (DL-SHG pro, Toptica) for resonant excitation of the $PbV^-$ centers. PSB is detected using a 561 nm long-pass filter for the PLE spectroscopy measurements. The wavelength during the PLE measurements is monitored by a wavelength meter (WS8-30, HighFinesse). The confocal fluorescence mappings and the PLE spectra are recorded by the Qudi module [81]. The $PbV^-$ centers are excited by a pulsed picosecond laser (D-TA-530B, PicoQuant) for the lifetime measurements. To observe the indistinguishability of photons from a single $PbV^-$ center in an SIL, we perform HOM interference experiments, for which avalanche photodiodes with lower dark count rates (SPCM-AQRH-16, Excelitas) are used. The details of the HOM experiment are described in Section 2.4. The HBT, HOM, and lifetime measurements are performed using a fast counter (Time Tagger Ultra, Swabian Instruments).

The Python codes used for making the figures was developed with the assistance of Chat GPT (Codex) and Gemini. All AI-assisted codes were reviewed, tested, and validated by the authors prior to use, and the authors take full responsibility for its accuracy and reproducibility.

**Acknowledgements**

We would like to thank Ryotaro Abe for experimental support. We also thank Materials Analysis Division, Core Facility Center, Institute of Science Tokyo for technical assistance. This work was supported by JSPS KAKENHI (Grants No. 25K24501), the MEXT Quantum Leap Flagship Program (Grant No. JPMXS0118067395), JST Moonshot R&D (Grant No. JPMJMS2062), JST ASPIRE JPMJAP24C1nd Council for Science, Technology and Innovation (CSTI), 3rd Cross-ministerial Strategic Innovation Promotion Program (SIP) Quantum.

# Supporting Information

for Enhanced Emission and Two-Photon Interference of Lead-Vacancy Centers in Diamond Solid Immersion Lenses

Koyo Hirai [1, †], Eiki Ota [1, †], Yiyang Chen [1], Ren Kato [1], Peng Wang [1, ‡], Toshiharu Makino [2], Takashi Taniguchi [3], Masashi Miyakawa [3], Shinobu Onoda [4], Mutsuko Hatano [1], and Takayuki Iwasaki [1, *]

[1]*Department of Electrical and Electronic Engineering, School of Engineering, Institute of Science Tokyo, Meguro, 152-8552 Tokyo, Japan*

[2]*Advanced Power Electronics Research Center, National Institute of Advanced Industrial Science and Technology, Tsukuba, 305-8568 Ibaraki, Japan*

[3]*Research Center for Materials Nanoarchitectonics, National Institute for Materials Science, Tsukuba, 305-0044 Ibaraki, Japan*

[4]*Takasaki Advanced Radiation Research Institute, National Institutes for Quantum Science and Technology, Takasaki, 370-1292 Gunma, Japan*

*Corresponding author: Takayuki Iwasaki (iwasaki.t.c5b4@m.isct.ac.jp).

[†]These authors contributed equally to this work.

[‡]Present address of Peng Wang: Institute for Quantum Optics, Ulm University, Albert-Einstein-Allee 11, D-89081 Ulm, Germany.

**SECTION S1 | AFM observation**

Figure S1 shows AFM images of the two samples (Sample 1 and 2). The regions close to the SILs are selected for analyses of surface roughness, marked with black rectangles. It should be noted that diamond samples subjected to HPHT treatment exhibit strong spatial variations [1] in surface morphology, resulting in large fluctuations in the surface roughness. Therefore, the present analysis is limited to the region surrounding SILs, and the possibility of even larger surface roughness in other regions cannot be excluded.

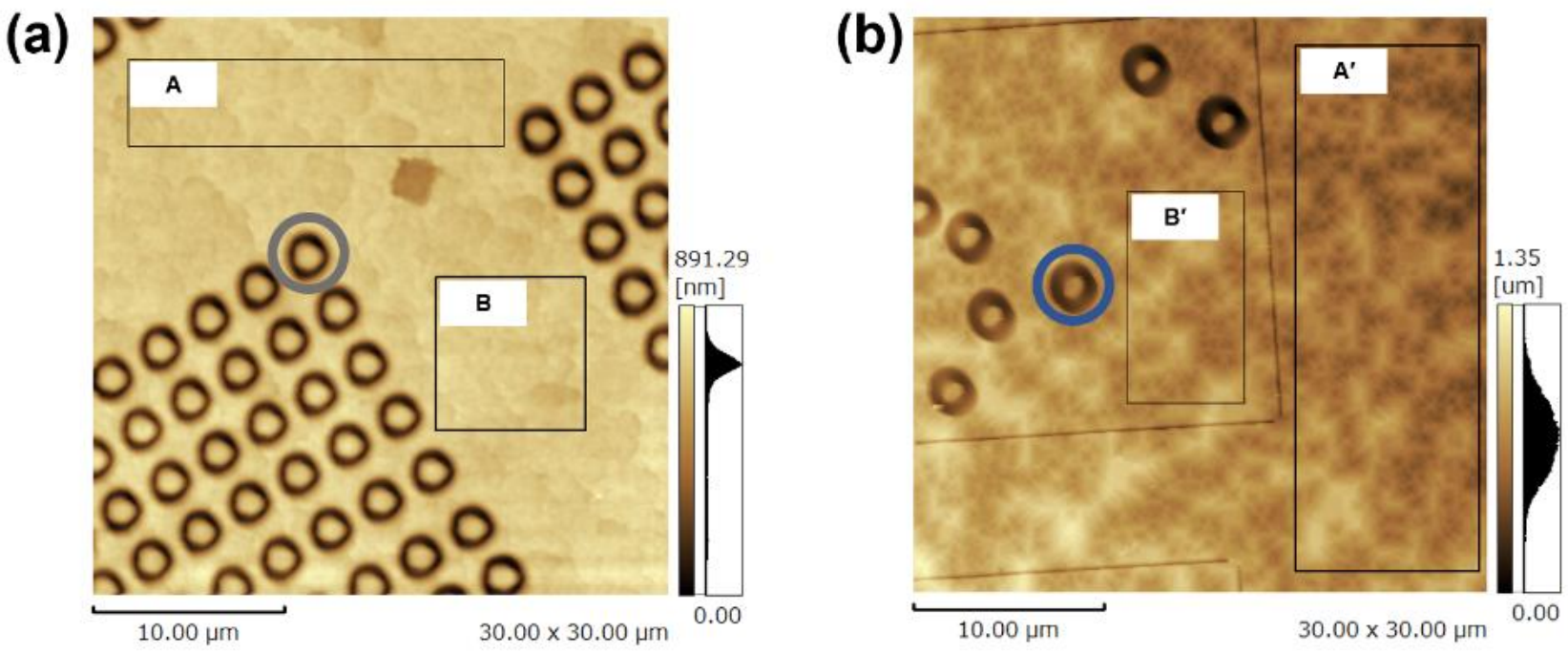

**Figure S1 |** AFM observation. (a) Sample 1. The SIL, enclosed by a gray circle, corresponds to a representative SIL shown in the magnified 3D-AFM (Fig. 1(e) bottom in the main text). (b) Sample 2. The SIL, enclosed by a blue circle is used for the HOM experiment (Section 2.4 in the main text).

## SECTION S2 | Top-view SEM images

Figure S2 shows top-view SEM images of SILs. An SIL in Sample 1 (Fig. S2(a)) exhibit a nearly round shape, whereas that in Sample 2 (Fig. S2(b)) becomes slightly elliptical with a major axis of ~1100 nm and a minor axis of ~890 nm. This is thought to be caused by the initial surface roughness before the FIB fabrication and FIB processing conditions including the focus adjustment.

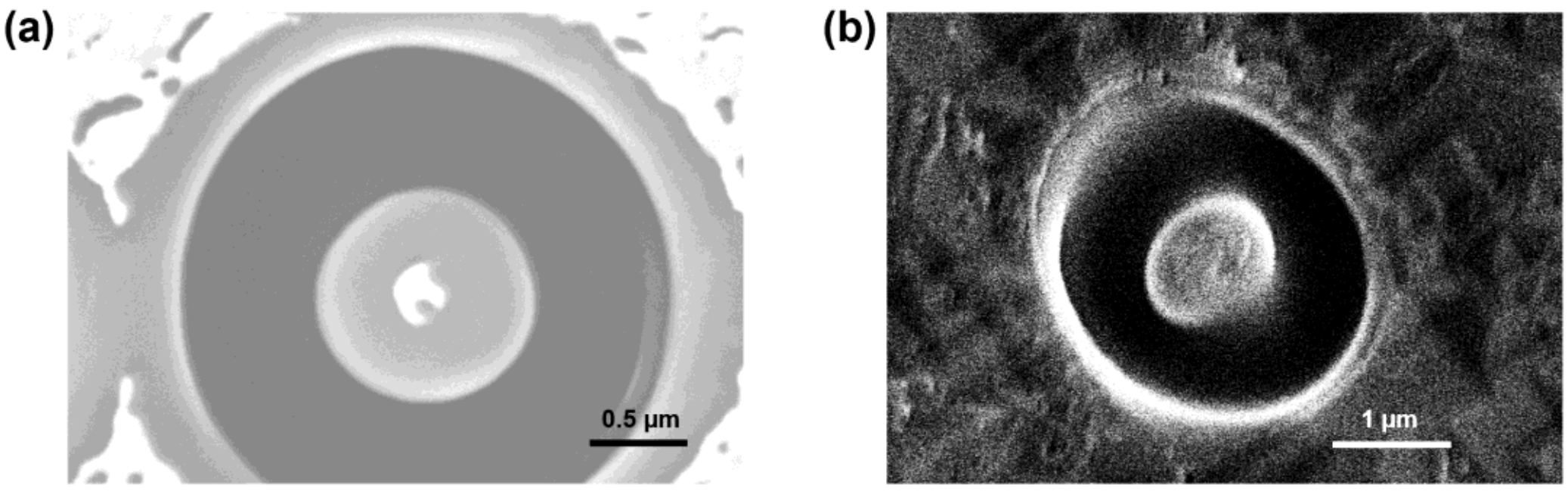

**Figure S2 |** Top-view SEM images of SILs in (a) Sample 1 and (b) Sample 2. The SIL shown in (a) is another SIL to that shown in Fig. 1(d).

## SECTION S3 | SIL fabrication on large pyramid-like surface structures

We try to fabricate SIL arrays on the surface with large pyramid-like structures formed during HPHT anneal (Fig. S3(a)). Note that this sample is another sample from Sample 1 and 2. Pyramid-like structures occasionally appear on the diamond surface as a result of diamond regrowth [1]. Some of SILs are fabricated on the pyramid-like structures and exhibit distorted morphologies (Fig. S3(b)). Since implanted Pb atoms are located below the original diamond surface, these distorted SILs should not contain $PbV^{-}$ centers.

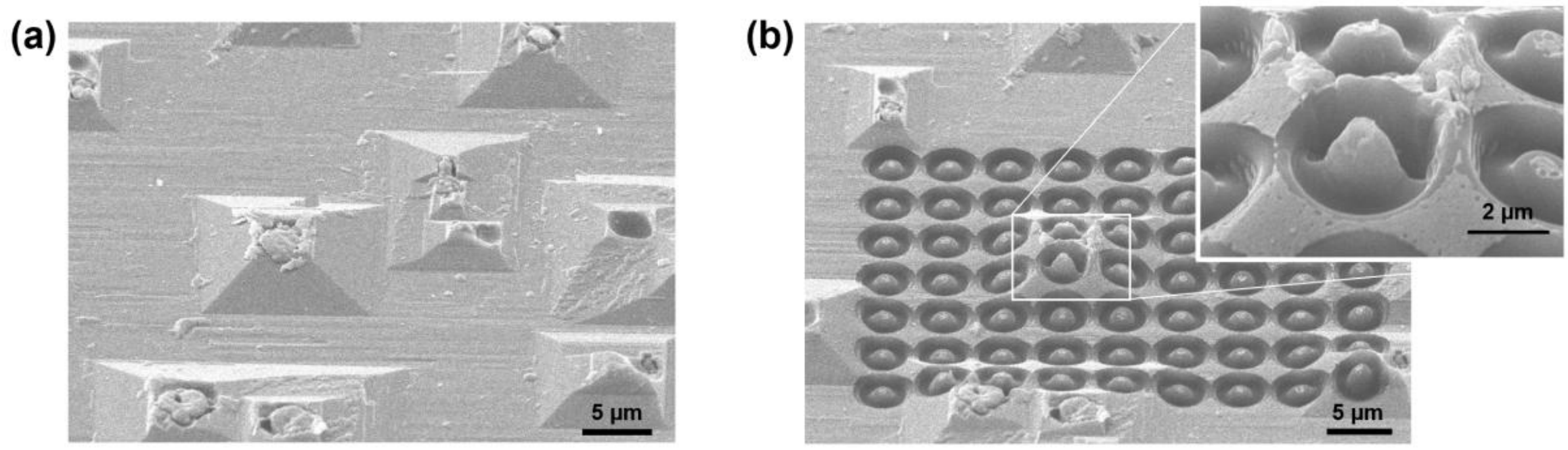

**Figure S3 |** SIL array on pyramid-like diamond structures. (a) SEM image before the SIL array fabrication. (b) SEM image of the SIL array fabricated on the large structures shown in (a). Inset: Magnified SEM image of SILs.

## SECTION S4 | Ensemble $PbV^{-}$ centers of bulk region in Sample 1

We investigate optical properties of bulk region in Sample 1. We select the region near SIL-1 (Fig. S4(a)). Based on CFM images, the count rate of the spot where we conduct PLE measurement is approximately one-quarter of SIL-1 (Figs. S4(a, b)). Figure S4(c) shows a PLE scan under 1.3 nW resonant excitation. We clearly observe multiple peaks, indicating that this spot consists of an ensemble of emitters. The larger number of peaks compared to the spectrum obtained for SIL-1 (Fig. 2(c) in the main text) is considered to arise from the higher excitation power and spatial variations. The higher excitation power is likely to excite emitters shifting from the center of the SIL. Figure S4(d) shows lifetime measurement of this spot, and the value of the lifetime is comparable to the emitters in SIL (Fig. 3(b) in the main text).

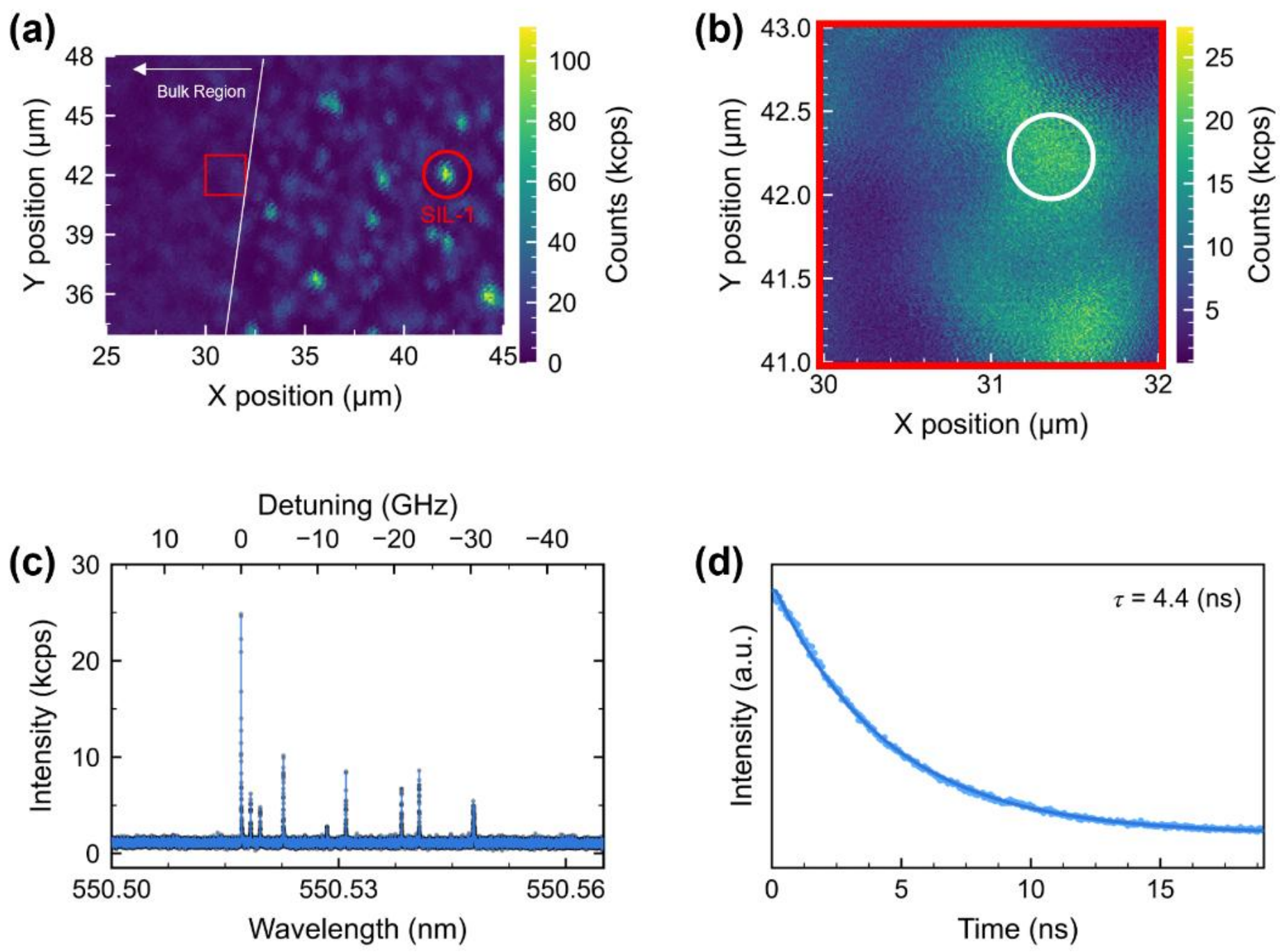

**Figure S4 |** Optical properties of bulk region in Sample 1. (a) CFM image of bulk region near SIL-1. The red square indicates the spot we select for PLE and lifetime measurements. (b) Magnified CFM image of the bulk area shown in (a). The target spot is marked with a white circle. (c) PLE spectrum of the target spot under 1.3 nW resonant excitation. (d) Lifetime measurement of the bulk spot.

## SECTION S5 | Two dominant emitters in SIL

As seen in Figs. 2(c, d) in the main text, one dominant peak with a high fluorescence intensity leads to the observation of single-photon like nature. We also observe two enhanced peaks with comparable intensities under resonant excitation in another SIL (Fig. S5(a)). The spot shows $g^2(0)$ of 0.57 in the auto-correlation function (Fig. S5(b)). This value corresponds to the presence of double emitters in the spot. Note that while the shorter-wavelength peak has a linewidth of ~40 MHz (Fig. S5(c)), another peak shows sudden drop of the intensity due to the charge state transfer (Fig. S5(d)) [2]. Thus, this peak is not included in the statistical analyses of the PLE spectra in the main text.

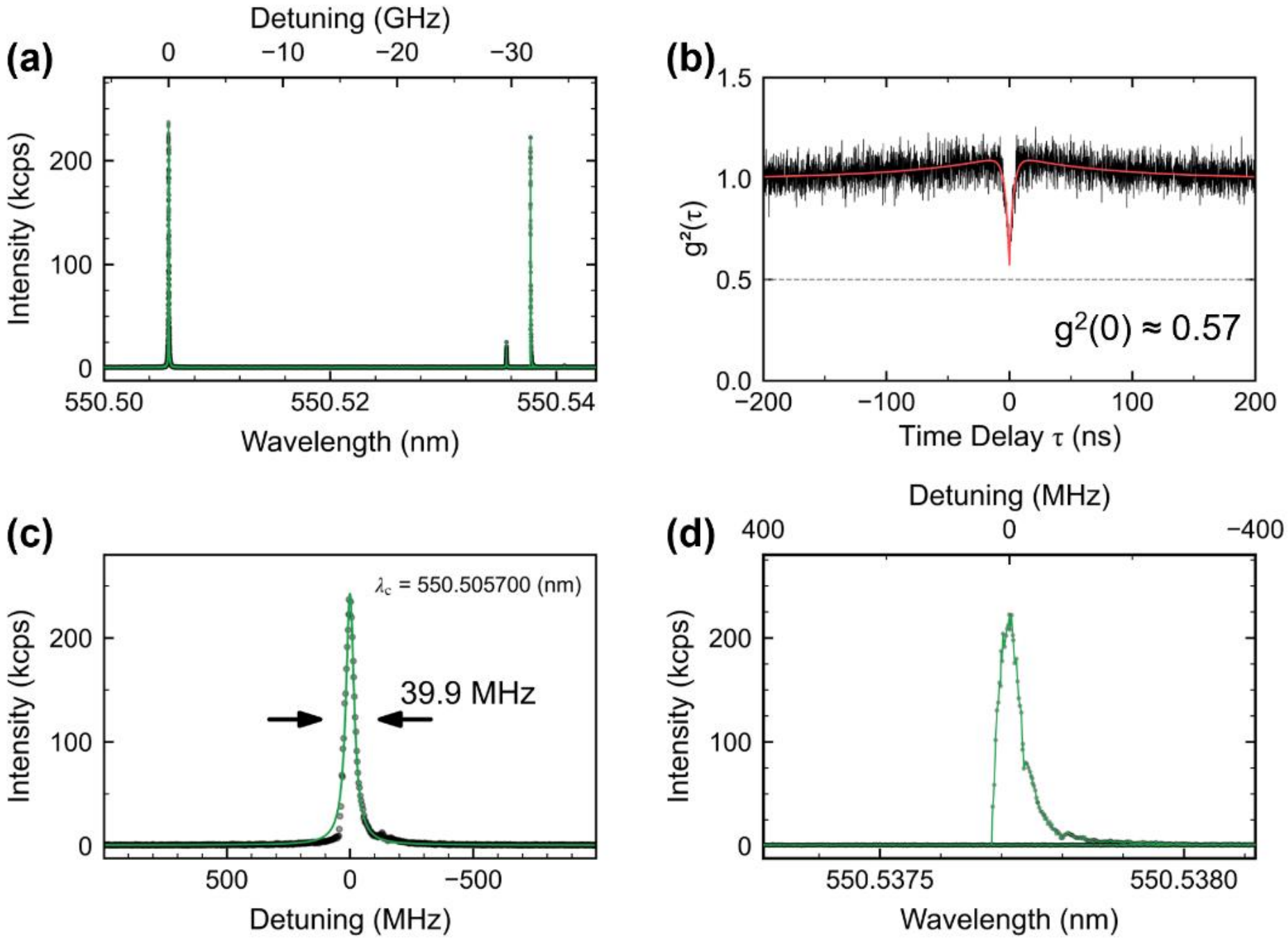

**Figure S5 |** Two enhanced $PbV^-$ centers in SIL. (a) PLE spectrum of the bright SIL under 0.6 nW resonant excitation. (b) HBT measurement. (c) The shorter-wavelength peak with a Lorentzian fit (FWHM = 39.9 MHz). (d) The longer-wavelength peak with charge state transfer.

## SECTION S6 | $PbV^-$ center in bulk diamond

An emitter in bulk diamond is shown in the saturation curve in Fig. 2(e) in the main text for comparison. Here, we show further optical characteristics of this emitter in Sample 2 (Fig. S6). It should be noted that Sample 2 discussed here is after the first annealing step (HPHT annealing time: 20 min). We observe unstable PLE peaks on several emitters after this first anneal step, while the $PbV^-$ center used as a reference in the saturation curve (Fig. 2(e)) is carefully characterized and is confirmed to have stable optical properties.

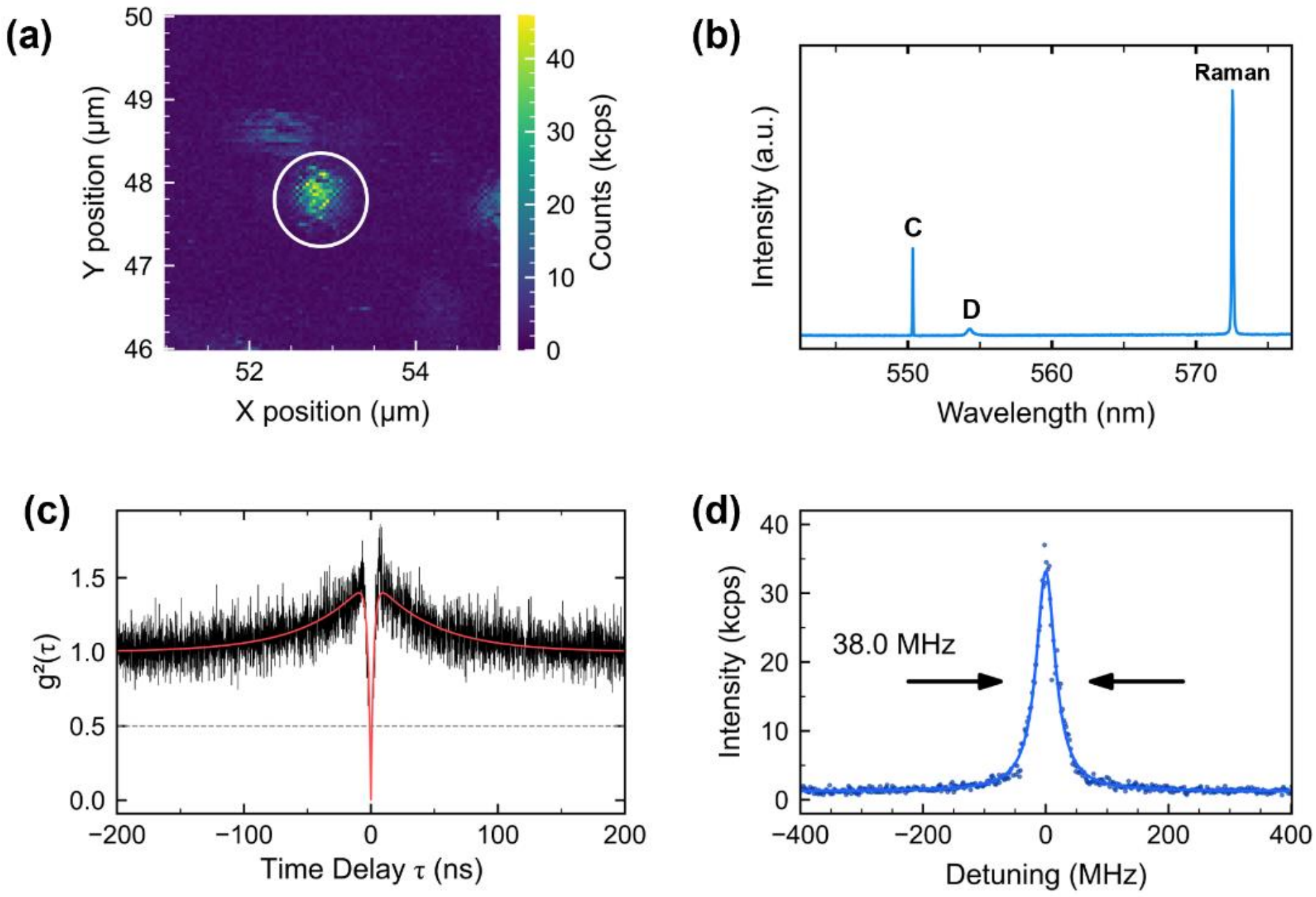

**Figure S6 |** Single $PbV^-$ center in bulk diamond. (a) CFM image of Bulk under 4 mW non-resonant excitation. (b) PL spectrum. (c) HBT curve. (d) PLE spectrum under 1.1 nW resonant excitation.

## SECTION S7 | PLE scan without soft ICP etching

We can observe clear PLE spectrum from SILs only after the soft ICP etching process [3], conducted after the FIB milling. Without the soft ICP etching, we observe a PLE peak only under simultaneous irradiation of the resonant and non-resonant lasers (Fig. S7). Surface damages induced during the FIB processing causes the charge instability of the $PbV^{-}$ center in SIL. The 532 nm non-resonant laser stabilizes the negative charge state of the PbV center [4], leading to the observation of the peak. Then, the excitation is immediately switched to resonant excitation only. The 532 nm non-resonant laser is turned off, and the power of the tunable resonant laser is adjusted using an acousto-optic modulator. The scan is performed again toward the observed resonance peak. However, we do not observe any PLE peak. Thus, we need to remove the Ga atoms and induced damages in SILs.

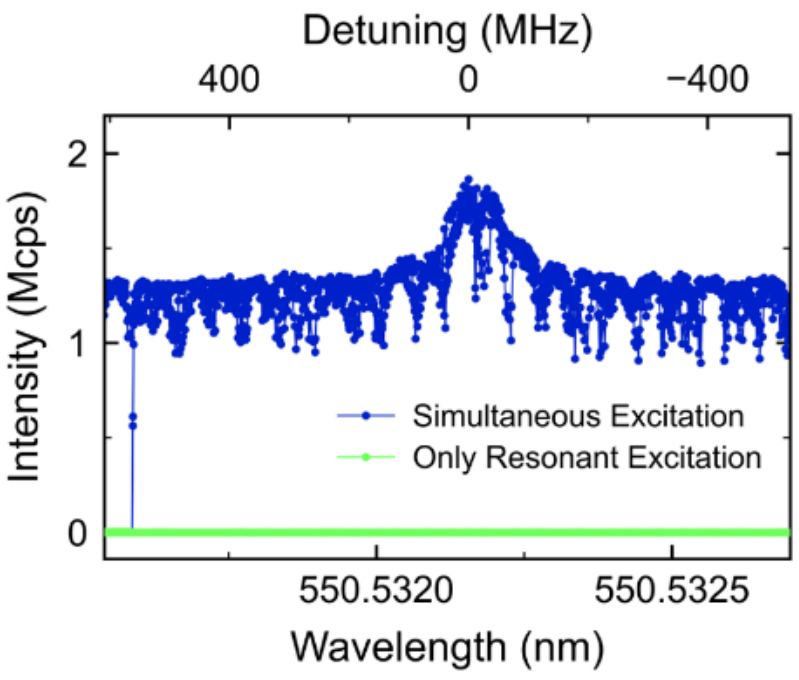

**Figure S7 |** PLE measurement in SIL before soft ICP etching. Simultaneous irradiation of non-resonant 532 nm laser (power: 200 μW) and a tunable resonant laser (power: 10 nW). The excitation power under resonant excitation only is 1.6 nW.

## SECTION S8 | PLE measurements in multiple SIL arrays

We show PLE peaks from 18 $PbV^-$ centers in different SIL arrays in Sample 1, used for the inhomogeneous and linewidth distributions in the main text. Other two emitters are in SIL-1 in Fig. 3(a) and in the SIL in Fig. S5(c).

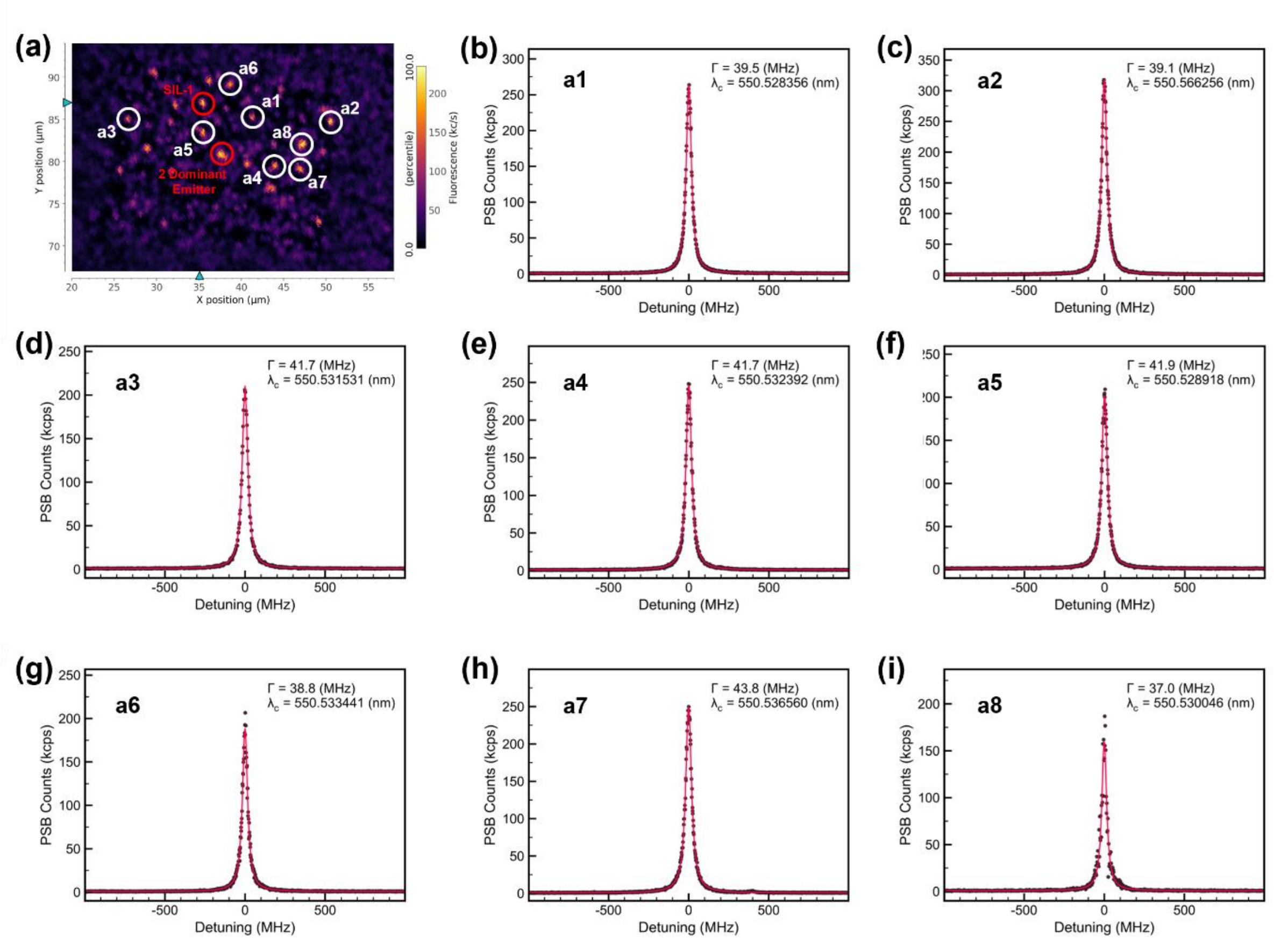

**Figure S8 |** Array-A. (a) CFM image. (b–i) PLE spectra.

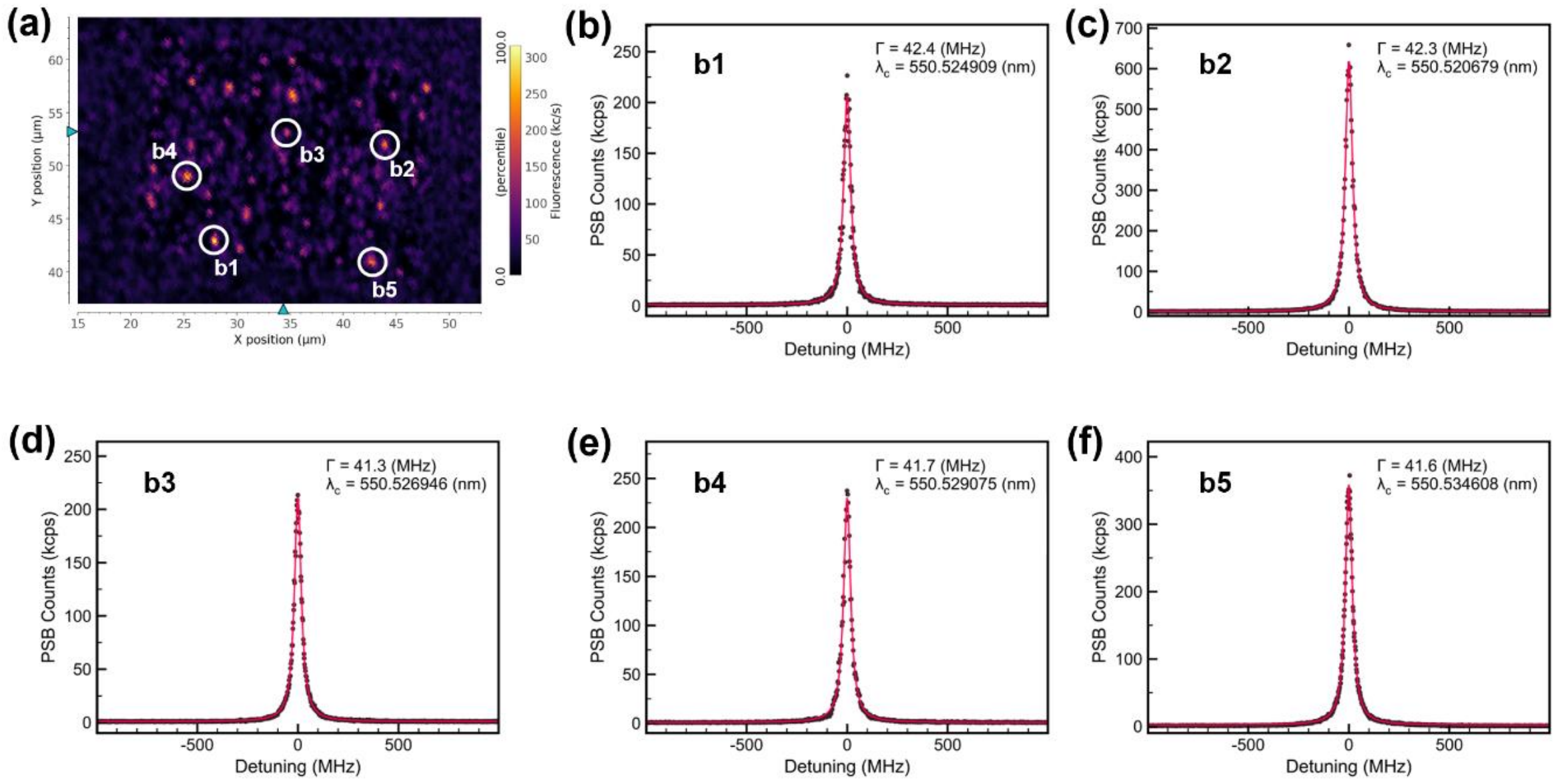

**Figure S9 |** Array-B. (a) CFM image. (b–f) PLE spectra.

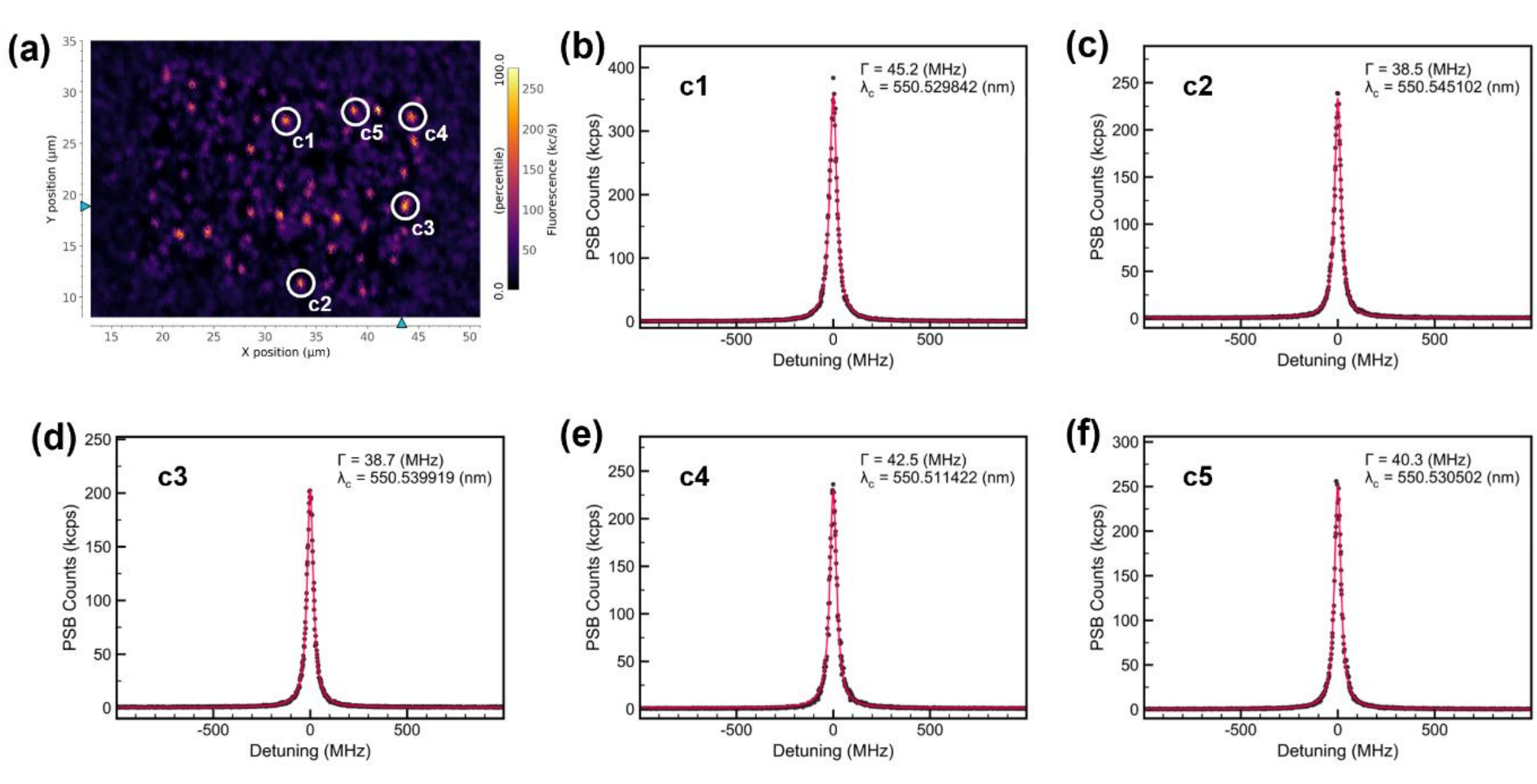

**Figure S10 |** Array-C. (a) CFM image. (b–f) PLE spectra.

## SECTION S9 | HOM interference

Figure S11 shows HOM interference in a wider delay time range. The second-order intensity correlation $g^2_{\mathrm{HOM}}(\tau)$ exhibits a three-dips, consisting of a central quantum interference dip at $\tau = 0$ and two symmetric side dips at $\tau = \pm\Delta\tau \sim \pm 44$ ns, where $\Delta\tau$ corresponds to the delay introduced by the difference in single-mode fiber length between the two arms. The two side dips originate from the intrinsic single photon anti-bunching (self-correlation) of consecutive photons [5, 6]. Crucially, the 44 ns optical delay is substantially longer than both the emitter lifetime. As a result, the side dips are well-isolated from the central main dip, ensuring that their tails do not overlap with or artificially shallow the minimum of the interference dip $g_{\parallel}(0)$.

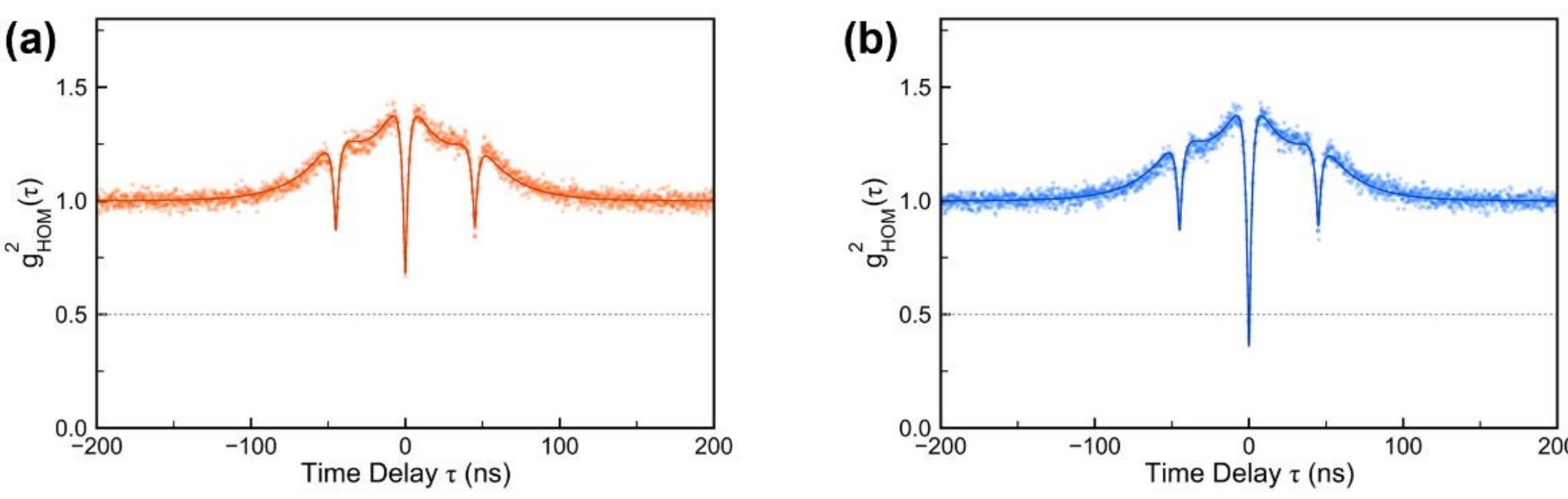

**Figure S11 |** HOM interference for (a) orthogonal polarization and (b) parallel polarization.

## SECTION S10 | SRIM simulation for Pb ions distribution

Figure S12 shows SRIM [7] simulation on Pb ion implantation at an acceleration energy of 4.5 MeV. We fit the simulation result with an equation [8]:

$$n(z) = \frac{\phi}{\sigma_p\sqrt{2\pi}} \exp\left[-\frac{\left(z - R_p\right)^2}{2\sigma_p^2}\right].$$

where fluence $\phi$, mean projected depth $R_p$, and straggle $\sigma_p$ are fitting parameters. From the fitting, we obtain the mean projected depth $R_p$ = 515.37 ± 0.36 nm and the straggle $\sigma_p$ = 52.03 ± 0.36 nm. On the basis of this information, we set the target height of SILs to be 500 nm.

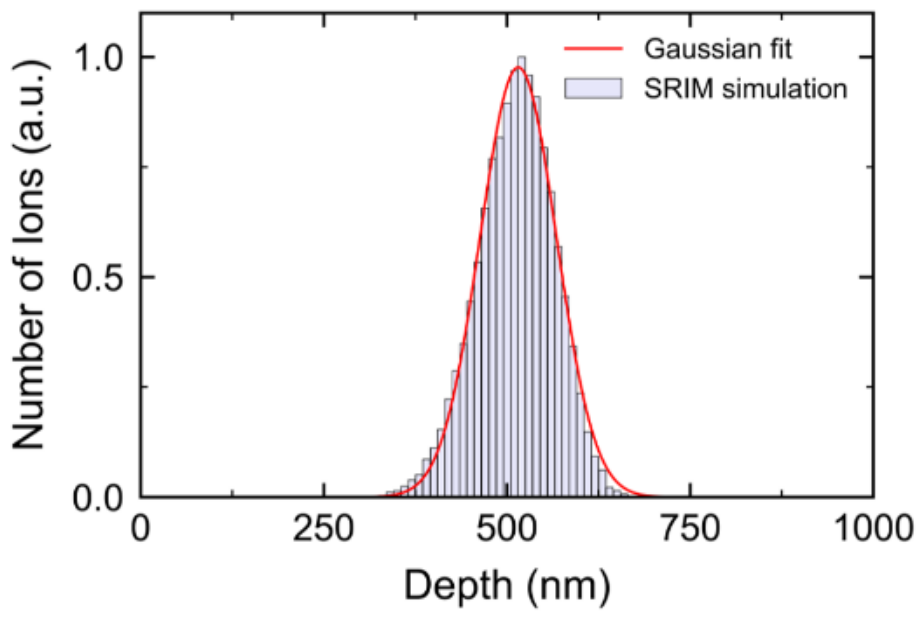

**Figure S12 |** SRIM simulation of Pb ion implantations into a diamond substrate at 4.5 MeV.

**References in Supporting Information**